\documentclass{amsart}
\usepackage{geometry}                
\usepackage[parfill]{parskip}    
\usepackage{graphicx}
\usepackage{amssymb}
\usepackage{amsmath}
\usepackage{bm}
\newtheorem{theorem}{Theorem}
\newtheorem{lemma}[theorem]{Lemma}

\usepackage{setspace}

\newcommand{\sE}{\mathcal{E}}
\newcommand{\sL}{\mathcal{L}}
\newcommand{\sZ}{\mathcal{Z}}
\newcommand{\PP}{{\mathbb P}}

\newcommand{\Vint}{\mathring{V}} 
\newcommand{\Eint}{\mathring{E}}

\newcommand{\be}{\begin{equation}}
\newcommand{\ee}{\end{equation}}

\title{Computing the Distribution of a Tree Metric}
\author{David Bryant and Mike Steel}
\thanks{We thank the Alexander von Humboldt Foundation and the Marsden Fund (DB) for supporting this work}

\address{DB: Mathematics Department, University of Auckland; MS: Department of Mathematics and
 Statistics, University of Canterbury, Christchurch, New Zealand}

\email{d.bryant@auckland.ac.nz, m.steel@math.canterbury.ac.nz}



\begin{document}

\begin{abstract}
The Robinson-Foulds (RF) distance is by far the most widely used measure of dissimilarity between trees.  Although the distribution of these distances has been investigated for
twenty years, an algorithm that is explicitly polynomial time has yet to be described for computing this distribution (which is also the distribution of trees around a given tree under the popular Robinson-Foulds metric).  In this paper we derive 
a polynomial-time algorithm for this distribution. We show how the distribution can be approximated by a Poisson distribution determined by the proportion of leaves that lie in `cherries' of the given tree. We also describe how our results can be used to derive normalization constants that are required in a recently-proposed maximum likelihood approach to supertree construction.
\end{abstract}
\maketitle
\section{Introduction}

Tree comparison metrics are widely used in phylogenetics for comparing evolutionary trees \cite{fels, Semple03}  and for performing statitistical tests - for example, to test whether two trees are `significantly different' from each other than one might expect if one or both trees were randomly chosen \cite{pen, pen2}.  In order to address these statistical questions one needs to determine the distribution of the metric under some null model (see, for example, \cite{pen, pen2}). The  {\em symmetric difference} or {\em Robinson-Foulds} metric is the most widely used measure of differences between phylogenetic trees, and its distribution is particularly attractive to study.  In a landmark paper \cite{Hendy84}, the authors described this distribution of trees relative to a fixed reference tree via a system of generating functions. This allowed the authors to calculate the distribution explicitly for small trees and provided a tool for analytic results on this distribution in later work by others.

However, the approach described in \cite{Hendy84} does not immediately appear to provide a polynomial-time algorithm for computing this distribution, and for larger trees their approach may be computationally prohibitive.
In this paper, we describe how to calculate the distribution of the Robinson-Foulds metric relative to a fixed tree. We also show how the distribution can be approximated by a Poisson distribution whose parameter depends on just one aspect of tree shape - the number of `cherries'.

Our investigation into the distribution of the metric has also been motivated by its relevance to a recent approach for `supertree' construction that is based on maximum likelihood \cite{Steel08}.  In particular, our algorithm allows the normalization constants in the likelihood calculations to be computed explicitly. We describe how these normalization constants depend weakly on aspects of the shape of the tree - for example, how many `cherries' the tree has.   We start by recalling some terminology.

Let $X$ be a finite set. A {\em phylogenetic tree} with leaf set $X$ is a tree with its degree one vertices (leaves) labelled bijectively by elements of $X$ and whose remaining vertices have degree at least three. We use $V(T)$ and $E(T)$ to denote the set of nodes (vertices) and edges of $T$. Let $\Vint(T)$ denote the set of internal (non-leaf) nodes of $T$ and let $\Eint(T)$ be the set of edges in $E(T)$ that have both endpoints in $\Vint(T)$, the internal edges.

A phylogenetic tree is {\em fully resolved} if every internal vertex has degree three. Following \cite{Hendy84} we let $PT(n)$ denote the set of phylogenetic trees on the finite set $X = \{1,2,\ldots,n\}$ and $BPT(n)$ the set of fully resolved (`binary') trees in $PT(n)$ (two trees in $BPT(6)$ are shown in Fig. 1).  The number of trees in $BPT(n)$ is denoted $b(n)$ and is given by:
\be b(n) = (2n-5)!! = \prod_{k=3}^n (2k-5)\,\hspace{2cm} n \geq 3,\ee
see \cite{Semple03}. For convenience, we let $\beta(m)$ denote the number of fully resolved trees with exactly $m$ internal edges, so:
\be
\beta(m) = b(m+3) = \prod_{k=3}^{m+3} (2k-5)\, \hspace{2cm}  m \geq 0.
\ee

Every edge $e \in E(T)$ induces a bipartition or {\em split} of the leaf set $X$ corresponding to the labels present in the two connected components remaining when the edge $e$ is removed. Let $\pi(T,e)$ denote this bipartition, which we consider unordered. We let $c(T)$ denote the set of all bipartitions obtained by removing different edges of $e$. Hence $|c(T)| \leq 2n-3$, the maximum number of edges in a phylogenetic tree, and $|c(T)| = 2n-3$ exactly when $T$ is fully resolved. A bipartition is {\em trivial} if it separates a single element from all other elements; trivial bipartitions correspond to the edges in the tree that are {\em external}, meaning that they are incident with a leaf of the tree. A {\em cherry} of a fully resolved phylogenetic tree $T$ is a pair of leaves that forms one half of a split of $T$ (i.e. a pair of leaves whose incident edges contain a common vertex). In Fig. 1 the pairs $(1,2)$ and $(5,6)$ form cherries in both trees, while the right-hand tree has an additional cherry $(3,4)$.

The {\em symmetric difference metric} is defined on $PT(n)$, and hence on $BPT(n)$, by:
\[d(T_1,T_2) = | c(T_1) \bigtriangleup c(T_2) |.\]
Note that this number is always even when $T_1$ and $T_2$ are both in $BPT(n)$, since, for any two trees in $PT(n)$, we have
$d(T_1, T_2) = |c(T_1)|+|c(T_2)|- 2|c(T_1) \cap c(T_2)|$, and if $T_1, T_2 \in BPT(n)$ then $|c(T_1)|=|c(T_2)|= 2n-3$. As an example,  the two trees shown in Fig. 1 have a distance value of $2$ since the splits $\{1,2,3\}|\{4,5,6\}$ and $\{3,4\}|\{1,2,5,6\}$ each occur in just one tree.

The metric was introduced by Bourque \cite{Bourque78} and generalised by Robinson and Foulds \cite{Robinson81}. As all phylogenetic trees contain all trivial splits, the maximum possible distance between two trees is $2(n-3)$, which is twice the maximum number of internal edges.

\section{Computing the distribution of the Robinson-Foulds metric}

For each $T \in PT(n)$, let $b_m(T)$ denote the number of trees $T' \in BPT(n)$ for which $d(T,T') = m$. As $d$ is a metric, $b_0(T) = 1$. A recursive formula for the generating function of $b_m(T)$ is given in \cite{Hendy84} and \cite{Steel88}. As far as we could deduce, the formula does not provide a polynomial time algorithm  for computing the $b_m(T)$ values,  due to an exponential explosion in the number of subcases.

Instead we use an alternative approach, applying results of \cite{Steel88}. Let $q_s(T)$ denote the number of trees in $BPT(n)$ that share exactly $s$ internal splits with $T$. Then for all $m=0,2,4, \ldots, 2(n-3),$ we have:
\be b_m(T) = q_{n-3-m/2}(T). \label{eq:bmqm} \ee
Define the polynomial
\be q(T,x) = \sum_{s=0}^{n-3} q_s(T) x^m. \ee

Let $E \subset \Eint(T)$ denote a subset of the set of internal edges of $T$. The forest $T-E$ has exactly $|E|+1$ components  $F_1,F_2,\ldots,F_{|E|+1}$. We use $\Eint(F_i)$ as a short-hand for the edges of $\Eint(T)$ that are contained in $F_i$.

Define
\be N_E(T) = \prod_{i=1}^{|E|+1} \beta(|\Eint(F_i)|) \label{eq:NET} \ee
Then $N_E(T)$  equals the quantity $\langle \Phi(E) \rangle$ defined in \cite{Steel88} (here assuming that $T$ is fully resolved). For $s\geq 0$ define
\[r_s(T) = \sum_{\substack{E \subseteq \Eint(T)\\|E|=s}} N_E(T),\]
the sum of $N_E$ over all subsets $E \subset \Eint(T)$ of cardinality $s$. For example, $r_0(T)$ equals $\beta(|\Eint(T)|) = \beta(n-3)$.
It was shown in \cite{Steel88} that the generating function
\[R(T,x) = \sum_{s \geq 0}  r_s(T) x^s\]
satisfies the identity
\be q(T,x) = R(T,x-1) . \label{eq:qRconnection} \ee
In what follows we derive a formula to evaluate the coefficients $r_s(T)$ so that we can compute the coefficients $b_m(T)$ via \eqref{eq:bmqm} and \eqref{eq:qRconnection}.

As usual, the computation applies dynamic programming. Let $v_0$ be the node adjacent to leaf $n$. Delete leaf $n$ and make $v_0$ the root of the tree, so that now every internal node has exactly two children. For each internal node $v$ let $T_v$ denote the subtree of $T$ containing $v$ and all of its descendants. Given a subset $E \subseteq \Eint(T_v)$, we define $N_E(T_v)$ as in \eqref{eq:NET}, where $F_1,\ldots,F_{|E|+1}$ will now be components of $T_v - E$ instead of $T-E$. We let $\kappa(v,E)$ denote the number of edges in the component of $T_v-E$ containing $v$. For $s,k \geq 0$, we let $\sE(v,s,k)$ denote the set of all subsets $E \subseteq \Eint(T_v)$ such that $|E| = s$ and $\kappa(v,E) = k$. Define
\be R(v,s,k) = \sum_{E \in \sE(v,s,k)}   N_E(T_v) \ee
so that if $v_0$ is the root of $T$ and  $s \geq 0$, we have:
\be r_s(T) = \sum_{k=0}^s R(v_0,s,k).\ee
We now derive a recursion for $R(v,s,k)$. As is customary, an empty summation equals zero.

\begin{lemma}
Suppose that $v \in \Vint(T)$. Then
\be
R(v,0,k) = \begin{cases}
		 \beta(k) & \mbox{ if $k= |\Eint(T_v)|$;} \\
		 0 & \mbox{ otherwise.}
		 \end{cases}
\ee
\end{lemma}

\begin{lemma}
Suppose that $s \geq 1$. For all $v \in \Eint(T)$ let $n_v = |\Eint(T_v)|$.
\begin{enumerate}
\item If $k > n_v$ then $R(v,s,k) = 0$.
\item  If $v \in \Vint(T)$ has no children in $\Vint$ and $s \geq 1$ then $R(v,s,k) = 0$.
\item If $v \in \Vint(T)$ has one child $v_1$ in $\Vint$ then
\be
R(v,s,k) = \begin{cases}
			\sum_{k_1\geq 0} R(v_1,s-1,k_1) & \mbox{ if $k=0$;} \\
			R(v_1,s,k-1) (2k+1)& \mbox{ otherwise;}
		\end{cases} \label{eq:vone}
\ee
\item If $v \in \Vint(T)$ has two children $v_1,v_2$ in $\Vint(T)$ then
\be
R(v,s,0) = \sum_{s_1=0}^{s-2} \left(\sum_{k_1 \geq 0} R(v_1,s_1,k_1) \right) \left(\sum_{k_2 \geq 0} R(v_2,s-2-s_1,k_2) \right).
\ee
\item If $v \in \Vint(T)$ has two children $v_1,v_2$ in $\Vint(T)$ and $k \geq 1$ then
\begin{eqnarray}
R(v,s,k) & =&  \sum_{s_1=0}^{s-1}  \left(\sum_{k_1 \geq 0} R(v_1,s_1,k_1) \right) R(v_2,s\!-\!1\!-\!s_1,k-1)  \beta(k) / \beta(k-1) \nonumber \\
&& +  \sum_{s_2=0}^{s-1}  \left(\sum_{k_2 \geq 0}  R(v_2,s_2,k_2) \right) R(v_1,s\!-\!1\!-\!s_2,k-1)  \beta(k) / \beta(k-1) \nonumber \\
&& + \sum_{s_1 = 0}^{s} \sum_{k_1 = 0}^{k-2} R(v_1,s_1,k_1) R(v_2,s\!-\!s_1,k\!-\!2\!-\!k_1) \frac{\beta(k)}{\beta(k_1)\beta(k\!-\!2\!-\!k_1)}. \nonumber \\ \label{eq:vtwo}
\end{eqnarray}
\end{enumerate}
\end{lemma}
\begin{proof}
Parts (1) and (2) follow from the definition of $R$. 
\begin{enumerate}
\item[(3)] Let  $e$ be the edge from $v_1$ to $v$. When $k=0$ it holds that  $E \in \sE(v,s,k)$ if and only if $E = E_1 \cup \{e\}$ for some $E_1 \in \sE(v_1,s-1,k_1)$, where $k_1$ ranges from $0$ to $s-1$. This gives the first case.  When $k \geq 1$, the edge $e$ connecting $v$ and $v_1$ is absent from every set in $\sE(v,s,k)$. Thus $E \in \sE(v,s,k)$ if and only if $E \in \sE(v_1,s,k-1)$.
\begin{eqnarray*}
N_E(T_v)& =&  N_{E'}(T_{v_1}) \frac{\beta(k)}{\beta(k-1)} \\
& = &  N_{E'}(T_{v_1})  (2k+1).
\end{eqnarray*}
\item[(4)] Let $e_1,e_2$ be the edges from $v$ to $v_1,v_2$ respectively. Since $k=0$, for all $E \in \sE(v,s,k)$, we have $e_1 \not \in E$ and $e_2 \not \in E$. Thus $E \in \sE(v,s,k)$ if and only if there exists $E_1 \in \sE(v_1,s_1,k_1)$ and $E_2 \in \sE(v_2,s-s_1,k_2)$ for some $s_1,k_1,k_2 \geq 0$ such that $E = E_1  \cap E_2$. For each such set $E$, we have: $N_E(T_v) = N_{E_1}(T_{v_1}) N_{E_2}(T_{v_2})$.
\item[(5)] Again, let $e_1,e_2$ be the edges from $v$ to $v_1,v_2$ respectively. For each $E \in \sE(v,s,k)$ with $k>0$, exactly one of the following cases holds:
\begin{enumerate}
\item[] {\em Case 1:} $e_1 \in E$ but $e_2 \not \in E$.
This case applies if and only there exists $E_1 \in \sE(v_1,s_1,k_1)$ and $E_2 \in \sE(v_2,s - 1 - s_1,k-1)$ for some $s_1,k_1 \geq 0$ such that $E = E_1 \cup E_2 \cup \{e_1\}$. For such a set $E$ we have
\[N_E(T_v)  =  N_{E_1}(T_{v_1}) N_{E_2}(T_{v_2}) \frac{\beta(k)}{\beta(k-1)}.\]
\item[] {\em Case 2:} $e_1 \not \in E$ but $e_2 \in E$.
Identical to Case 1 with $v _1$ and $v_2$ switched.
\item[]  {\em Case 3:} $e_1 \in E$ and $e_2 \in E$.
This case applies if and only there exists $E_1 \in \sE(v_1,s_1,k_1)$ and $E_2 \in \sE(v_2,s  - s_1,k-k_1 - 2)$ such that $E = E_1 \cup E_2 \cup \{e_1,e_2\}$. For each such set $E$ we have:  \[N_E(T_v) = N_{E_1}(T_{v_1}) N_{E_2}(T_{v_2}) \frac{\beta(k)}{\beta(k_1)\beta(k\!-\!2\!-\!k_1)}.\]
\end{enumerate}
\end{enumerate}
\end{proof}

\begin{theorem}
Given a fully resolved tree $T$ on $n$ leaves the coefficients $b_m(T)$ can be computed in $O(n^5)$  time.
\end{theorem}
\begin{proof}
	Consider a vertex $v \in \Vint(T)$. If $v$ has one child in $\Vint(T)$ then we evaluate \eqref{eq:vone} for all $s,k \leq n-3$ in $O(n^3)$ time. If $v$ has two children in $\Vint(T)$ then we evaluate \eqref{eq:vtwo} in $O(n^4)$ time.
	
	Hence computing all the coefficients $r_s(T)$ takes $O(n^5)$ time. From \eqref{eq:qRconnection}, we obtain:
	\be
	q_m(T) = \sum_{s=m}^{n-3} \binom{s}{m} r_s(T) (-1)^{s-m},
	\ee
	from which we compute the values $b_m(T) = q_{n-3-m/2}(T)$.
	\end{proof}

\section{Poisson approximation}

When $n$ is large we can approximate the $q_s(T)$ values by a Poisson distribution with mean $\lambda_T: = c_T/2n$ where
$c_T$ denotes the number of cherries of $T$ (recall that a {\em cherry} is a pair of leaves whose incident edges contain a common vertex). More precisely, we have the following result.
\begin{theorem}
\label{Poisson}
For any tree $T \in BPT(n)$, let $Y_T$ be a Poisson random variable with mean $\lambda_T$. Then the distributions $q_s(T)/b(n)$ (the proportion of trees in $BPT(n)$ that share $s$ nontrivial splits with $T$) and $Y_T$ have variational distance that converges to zero as $n \rightarrow \infty$. In particular,
$$\sum_{s\geq 0} |q_s(T)/b(n) - e^{-\lambda_T}\lambda_T^s/s!| = O(n^{-1}).$$
\end{theorem}

\begin{proof}
Let $X_T$ denote the random variable which counts the number of non-trivial splits that $T$ shares with a tree $T'$ selected uniformly at random from $BPT(n)$. Thus, $\PP(X_T = s) = q_s(T)/b(n)$. Let $X'_T$ be defined in the same ways as for $X_T$ but counting only splits that divide the leaf set into subsets of size $2$ and $n-2$. Clearly, $X'_T \leq X_T$. Moreover, the probability that $T'$ shares a split with $T$ that is not of the type counted by $X_T'$ is bounded above by a term of order $n^{-1}$ and so we have:
\begin{equation}
\label{lim1}
\PP(X_T \neq X'_T) = O(n^{-1}).
\end{equation}
Now, for any two discrete random variables $X$ and $X'$ an elementary probability argument shows that $\sum_s|\PP(X=s)-\PP(X'=s)| \leq 2\PP(X \neq X')$, and so:
\begin{equation}
\label{lim2}
\sum_{s\geq 0}|\PP(X_T=s) - \PP(X'_T=s)| \leq 2\PP(X_T \neq X'_T).
\end{equation}
Combining (\ref{lim1}) and (\ref{lim2}) gives:
\begin{equation}
\label{lim3}
\sum_{s\geq 0}|\PP(X_T=s) - \PP(X'_T=s)| = O(n^{-1}).
\end{equation}
By the triangle inequality,
\begin{equation}
\label{lim4}
\sum_{s\geq 0}|\PP(X_T=s) - \PP(Y_T=s)|\leq  \sum_{s\geq 0}|\PP(X_T=s) - \PP(X'_T=s)| + \sum_{s\geq 0}|\PP(X'_T=s) - \PP(Y_T=s)|
\end{equation}
which, combined with (\ref{lim3}), gives:
\begin{equation}
\label{lim5}
\sum_{s\geq 0}|\PP(X_T=s) - \PP(Y_T=s)|\leq  \sum_{s\geq 0}|\PP(X'_T=s) - \PP(Y_T=s)| +O(n^{-1}).
\end{equation}
Thus, to establish Theorem~\ref{Poisson} it suffices to show that
\begin{equation}
\label{lim6}
\sum_{s\geq 0}|\PP(X'_T=s) - \PP(Y_T=s)| = O(n^{-1}).
\end{equation}
Now, by Lemma 3 of \cite{Steel88}, we have:
\begin{equation}
\label{lim7}
\PP(X'_T = s) = \sum_{r=s}^{c_T} (-1)^{r+s}\binom{r}{s}\binom{c_T}{r}\frac{b(n-r)}{b(n)}.
\end{equation}
Furthermore, letting $\lambda$ denote $\lambda_T$ for brevity, we have:
$$\PP(Y_T=s) = e^{-\lambda}\lambda^s/s! = \sum_{r=s}^\infty (-1)^{r+s}\binom{r}{s}\frac{\lambda^r}{r!}.$$
Substituting this and (\ref{lim7}) into the left-hand side of (\ref{lim6}) gives the expression:
\begin{equation}
\label{lim8}
\sum_{s \geq 0}\left|\sum_{r=s}^\infty (-1)^{r+s}\binom{r}{s}\left[ \binom{c_T}{r} \frac{b(n-r)}{b(n)} - \frac{\lambda^r}{r!}\right]\right|
\end{equation}
which, after some algebra, and moving the absolute value inside the second summation, is bounded above by:
\begin{equation}
\label{lim9}
\Delta_n:= \sum_{s \geq 0} \frac{1}{s!}\sum_{r=s}^\infty \frac{1}{(r-s)!}f(n,r)
\end{equation}
where $$f(n,r) := \left(\frac{c_T}{2n}\right)^r \cdot \left|\frac{\prod_{i=1}^{r-1}(1-i/c_T)}{\prod_{j=1}^{r}(1-(2j+3)/2n)} - 1\right|$$
Using the fact that $c_T \leq n/2$, and a somewhat tedious case analysis, it can be shown that $f(n,r) \leq C/n$ for a constant $C$ that is independent of $r, n$.
It follows that  $$\Delta_n \leq   \sum_{s \geq 0} \frac{1}{s!}\sum_{r=s}^\infty \frac{1}{(r-s)!} C/n = Ce^2/n,$$ which establishes
(\ref{lim6})  and thereby the theorem.
\end{proof}

{\bf Remark}  If $T$ is selected uniformly at random from $BPT(n)$, then $\lambda_T$ converges in probability to $\frac{1}{8}$ (since the variance of $\lambda_T$ is $O(n^{-1})$ by Theorem 4(b) of \cite{mck}).  Thus, Theorem~\ref{Poisson} can be viewed as a refinement of the main result from \cite{Steel88} that for two trees selected uniformly at random from $BPT(n)$ the number of non-trivial splits they share is asymptotically Poisson distributed with mean $\frac{1}{8}$.

\section*{Application to Likelihood based supertrees}

Rodrigo and Steel \cite{Steel08} recently presented a likelihood framework for constructing consensus trees and supertrees. Let $\sL(T_i)$ denote the set of leaves of a (fully resolved) gene tree $T_i$. The probability of observing $T_i$ with leaf set $\sL(T_i)  = X_i$ given an estimated species tree or supertree $T$ has the form
\be  \PP_{T,X_i}(T_i) =  \PP_{T}(T_i)  = \frac{1}{\sZ_{T|\sL(T_i)}}  e^{-\beta_i d(T_i, T|\sL{T_i})} \ee
where $T|\sL(T_i)$ denotes the restriction of $T$ to the leaf set $T_i$, and where $\beta_i$ is a positive constant.
The normalising constant
\be
\sZ_{T_i}  = \sZ^i_T \,\, \,\,\,=  \hspace{-2mm} \sum_{T':\sL(T')=\sL(T_i)} e^{- \beta_i d(T',T|\sL{T_i})}
\ee
 is required so that the $\PP_{T}(T_i)$ values sum to $1$ over all choices of $T_i$.
 One complication with this approach is that the normalising functions $\sZ_{T_i}$ depend on  $T$ (more precisely, although $\sZ_{T_i}$ does not depend on how the leaves of $T$ are labeled, it may depend on the shape of $T$), meaning that the constant needs to be computed in order to compare the likelihood values of two trees.  This was overlooked in \cite{Steel08}, in particular Proposition 1 of that paper may only hold in certain cases (for example, if the sets $X_i$ are of size at most $5$, or if the $\beta_i$ values are sufficiently large). However, Proposition 1 of \cite{Steel08} can be corrected by replacing the term $$\sum_{i=1}^k \beta_i d(T_i, T|X_i)$$ in the statement of that Proposition by
$$\sum_{i=1}^k \beta_i d(T_i, T|X_i) + \gamma_i(T),$$ where $$\gamma_i(T) = \sum_{i=1}^k\log(\sZ_{T_i}) = \log(1+\sum_{m > 0} e^{-\beta_i m} n_m(T)),$$
and where $n_m(T)$ is the number of fully resolved phylogenetic trees on leaf set $X_i$ that have distance $m$ from $T|X_i$.

In general, normalising constants are difficult to evaluate.  When $d$ is the Robinson-Foulds distance, however, computing the constant is straight-forward. Suppose that $|X_i|=n$ and that $b_m(T)$ has been computed for all $m$. Then (suppressing the index $i$) we have:
\begin{eqnarray*}
\sZ_T & = & \sum_{T' \in BPT(n)}  e^{-\beta d(T,T') }  \\
& = & \sum_{m} b_m(T) e^{-\beta m } .
\end{eqnarray*}
which can be evaluated directly from the $b_m(T)$ values, and thereby in polynomial time overall in $n$.

It is instructive to estimate $\sZ_T$ in two limiting cases - firstly for values of $\beta$ that are close to $0$, and for values of $\beta$ that are large.
In both cases we find that the dominant aspect of the shape of $T$ affecting $\sZ_T$ is the number $c_T$ of cherries that $T$ has. The experimental performance of these approximations is evaluated in the final section. 

\subsection{Small values of $\beta$}

For $\beta$ close to $0$, we exploit the identity $e^{-\beta m} = 1 - \beta m +O(\beta^2)$ and write:
\begin{equation}
\label{sZ}
\sZ_T = b(n) -  \beta \sum_{m}mb_m(T) + O(\beta^2).
\end{equation}
Now, the first term in (\ref{sZ}) in $b(n)$ times the
expected RF distance (denoted $\nu(T)$) from $T$ to a tree that is slected uniformly at random from $BPT(n)$. From \cite{Steel88} (p.550), we have:
\be
\label{nunu}
\nu(T)/b(n) = 2n-6 - 2\sum_{i\geq 2} n_i \frac{b(i+1)b(n-i+1)}{b(n)},
\ee
where $n_i$ is the number of interior edges of $T$ for which the smaller subtree in $T-e$ contains $i$ leaves of $T$.
For example, consider the $105$ fully resolved trees with six leaves, each of which has one of two possible shapes, depending on whether it has two or three cherries (as shown in Fig. 1).
For any such tree $T_2$ with two cherries we have:
$$\nu(T_2)/b(n) = 6-\frac{26}{35},$$
while for any tree $T_3$ with three cherries we have:
$$\nu(T_3)/b(n) = 6-\frac{30}{35}.$$
Returning to the general setting,  we can expand (\ref{nunu}) and write:
\begin{equation}
\label{secondorder}
\nu(T) /b(n) = 2n-6 - 2\frac{c_T}{2n-5} - 6\frac{\tau_T}{(2n-5)(2n-7)} -O(n^{-2}),
\end{equation}
where $\tau_T$ is the number of edges $e$ of $T$ for which one of the subtree of $T-e$ has exactly three leaves of $T$.
Notice that: $$\tau_T \leq c_T,$$
since any $3$--leaf subtree necessarily contains a cherry; therefore a corollary of (\ref{secondorder}) is
\begin{equation}
\label{firstorder}
\nu(T) /b(n) = 2n-6 - 2\frac{c_T}{2n-5}  - O(n^{-1}),
\end{equation}
and so, from (\ref{sZ}), we have $$\sZ_T = b(n)\left(1- \beta (2n-6- 2\frac{c_T}{2n-5}  - O(n^{-1}))+ O(\beta^2)\right).$$
Thus, as $\beta$ converges to $0$, $\sZ_T$ converges to a constant, and when $\beta$ is close to  $0$, the small difference from this constant is dominated by $c_T$.

\subsection{Large values of $\beta$}

When $\beta$ is large,  let $\epsilon = e^{-2\beta}$. Then,
$$\sZ_T =1 + b_2(T)\epsilon + b_4(T)\epsilon^2 + O(\epsilon^3).$$
Now, $b_2(T) = 2(n-3)$, and from Theorem 2.26  of \cite{SteelThesis} we have:
$$b_4(T) = 4\binom{n-3}{2} + 6(n-6+c_T).$$
Thus if we let $A_{n, \epsilon} := 1+(2n-3)\epsilon+ 2(n^2-4n-6)\epsilon^2$ then
$$\sZ_T = A_{n, \epsilon} +6c_T\epsilon^2 + O(\epsilon^3).$$
Once again we see that in the limit (in this case, as $\beta$ tends to infinity) $\sZ_T$ converges to a constant, and for large values of $\beta$,
the small difference from this constant is dominated by $c_T$.

\section{Experimental results}

To study general features of the distribution, and examine the accuracy of the above approximations, we generated random trees and computed the distribution of the Robinson Foulds distance for each tree. The trees were drawn from a uniform distribution, with the number of taxa varying from 5 to 50. One thousand replicates were performed for each number of taxa. We also constructed an unrooted caterpillar tree and a balanced unrooted tree for every set of taxa. A balanced unrooted tree is one that minimises the length of the longest path between any two leaves, an example being the right-hand tree in Fig. 1. 

As predicted from the Poisson approximation, the distributions of Robinson-Foulds distances from a fixed tree were highly peaked. For all of the trees examined, at least $99\%$ of trees are either at distance $2(n-3)$, the maximum possible, or distance $2(n-4)$. 

For $T \in BPT(n)$, let $N_k(T)$ denote the number of trees in $BPT(n)$ within Robinson-Foulds distance $k$ of $T$: that is,
\[N_k(T) = \sum_{m=0}^k b_m(T).\]
Then $N_2(T) = 2(n-3)+1$, the number of trees that share all but one split with $T$, together with the tree $T$ itself. When $k > 2$, the value of $N_k(T)$ varies with the shape of $T$. We observed that for all $k$, $N_k(T)$ was minimised when $T$ is a caterpillar. At the other extreme, $N_k(T)$ was almost always maximised when $T$ was balanced, the exception being when $T$ was balanced but did not have the maximum number of cherries.

For each tree, and a range of different values for $\beta$, we computed the exact normalising constant $\sZ_T$. Fig. 2  illustrates the variation in $\sZ_T$ over different values of $\beta$, displayed on a log-log plot. The central curve gives the average $\sZ_T$ values for $1000$ fifty-taxa trees drawn from a uniform distribution, as a function of $\beta$. The small-$\beta$ and large-$\beta$ approximate values for $\sZ_T$ are also plotted.

As a function of $\beta$, the normalising constant has three distinct phases. For $0 \leq \beta < 0.03$ the normalising constant $\sZ_T$ is close to the total number of fully resolved trees, and is fit well by the small-$\beta$ approximation. For $\beta>3.0$ the normalising constant approaches $1$, and is close to the large-$\beta$ approximation. Between $0.03$ and $3$, the $\sZ_T$ value changes quickly as a function of $\beta$. In this interval, neither of the above approximations work well.

As we observed above, to correctly compute the likelihood for a supertree under the model of \cite{Steel08} we need to compute $\sZ_T$ for every distinct supertree $T$. Even though this calculation take polynomial time, it is still extremely expensive computationally, particularly considering that millions of candidate supertrees may be considered. We ask, then, the extent to which this computation is strictly necessary. In particular, if we ignore the normalising constant when comparing likelihoods, would the relative likelihood ordering of distinct trees change. The key question is then to determine how much the normalisation constants $\sZ_T$ vary. If the difference is sufficiently small then there will be no impact of ignoring the differences between normalising constants.

For a given value of $\beta$ define the {\em range} of $\sZ_T$ to be the ratio of the largest to the smallest $\sZ_T$ values over all fully-resolved trees with $n$ taxa.
Fig. 3 plots the range of $\sZ_T$ for the values of $\beta$ used in Fig. 2, and for $n=10,20,30,40,50$ taxa trees, on a log-log axis. The trees minimising $\sZ_T$ were always caterpillar trees and the trees maximising $\sZ_T$ were usually, but not always, balanced trees. The figure indicates that when $\beta$ is outside the range $[0.03,3]$ there is little variation in $\sZ_T$ between different trees. With $50$ taxa, the normalising constants differ by a maximum of 7.5 log-units. 

Suppose that we are comparing the log-likelihood of two trees $T_1$ and $T_2$ with respect to a third tree $T$. If $d_{RF}(T,T_1) \neq d_{RF}(T,T_2)$ then 
\[| \log( e^{-\beta d(T,T_1)}) - \log(e^{-\beta d(T,T_2)}) | \geq 2 \beta \]
so ignoring the normalising constant will only change the order of likelihood values if $| \log \sZ_{T_1}  -  \log \sZ_{T_2} | \geq 2 \beta$. Plotting the curve for $2 \beta$ on Fig. 3 we see that $| \log \sZ_{T_1}  -  \log \sZ_{T_2} | \geq 2\beta$ 
for some pairs of 50-taxa trees only when $\beta$ lies in the interval $[1.25,1.86]$. 
The corresponding interval will be even smaller for trees with fewer taxa: for 20 taxa trees there is no value of $\beta$ for which ignoring $\sZ_T$ scores leads to a switch in the order of likelihood values for two trees.

In summary, when $\beta$ is approximately $1.5$, and the number of taxa is greater than around $20$, it is potentially important to correctly compute normalisation constants. Outside that range, the influence of $\sZ_T$ on likelihood rankings can be safely ignored. 
We note, however, that here we are only interested in relative ordering of supertrees with respect to likelihood: a Bayesian Monte-Carlo approach may well need accurate $\sZ_T$ values  for all $\beta$.

\bibliographystyle{plain}

\begin{thebibliography}{1}\bibitem{Bourque78}M.~Bourque, ``Arbes de Steiner et reseaux dont varie l'emplagement de certains  sommets,'' PhD thesis, Universit\'{e} de Montr\'{e}al, Qu\'{e}bec, Canada, 1978.

\bibitem{fels} J. Felsenstein, {\em Inferring phylogenies}. Sinauer Press, 2004.

\bibitem{Hendy84}M.D. Hendy, C.H.C. Little, and D.~Penny, ``Comparing trees with pendant vertices labelled,'' {\em SIAM Journal of Applied Mathematics},  vol. 44, no. 5, pp. 1054--1065, 1984.

\bibitem{mck}
A. McKenzie and M. Steel, ``Distributions of cherries for two models of trees,'' {\em Mathematical Biosciences}, vol. 164, pp. 81--92, 2000.

\bibitem{pen}
D. Penny, M.A. Steel and E. Watson, ``Trees from languages and genes are very similar,'' {\em Systematic Biology}, vol. 42, no. 3, pp. 382--384, 1993.

\bibitem{pen2}
D. Penny, L.R. Founds, and M. D. Hendy, ``Testing the theory of evolution by Comparing phylogenetic trees constructed from five different protein sequences,'' {\em Nature}, vol. 297, 197--200, 1982.


\bibitem{Robinson81}D.F. Robinson and L.R. Foulds, ``Comparison of phylogenetic trees,'' {\em Mathematical Biosciences}, vol. 53, pp. 131--147, 1981.

\bibitem{Semple03}C.~Semple and M.~Steel, {\em Phylogenetics}. Oxford University Press, 2003.

\bibitem{Steel08}M.~Steel and A.~Rodrigo, ``Maximum likelihood supertrees,'' {\em Systematic Biology}, vol. 57, no. 2, pp. 243--250, 2008. 

\bibitem{SteelThesis} M.~Steel, ``Distributions on bicoloured evolutionary trees,'' PhD Thesis, Massey University, Palmerston North, New Zealand, 1989.

\bibitem{Steel88}M.~A. Steel, ``Distribution of the symmetric difference metric on phylogenetic trees,'' {\em SIAM J. Discrete Math.}, vol. 1, no. 4, pp. 541--551, 1988.

\bibitem{stepen} M.A. Steel and D. Penny,  ``Distributions of tree comparison metrics - some new results,'' {\em Systematic Biology}, vol. 42, no. 2, pp. 126--141, 1993.



\end{thebibliography}

\pagebreak
   \vspace*{1cm}

\begin{figure}[ht] \begin{center}
\resizebox{10.0cm}{!}{
   \includegraphics[width=10cm]{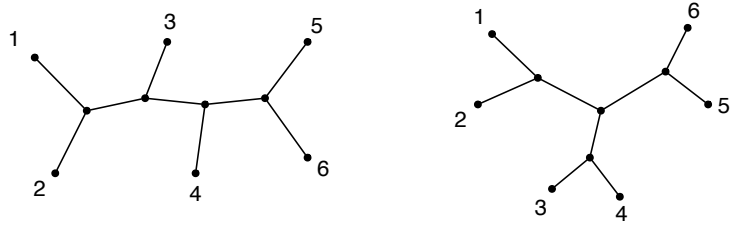}
}   \vspace{3cm}
\caption{Two fully resolved phylogenetic trees on six leaves, with Robinson-Foulds distance two.}
\end{center}
\label{figure1}
\end{figure}

\pagebreak
   \vspace*{1cm}

\begin{figure}[ht] \begin{center}
\resizebox{10.0cm}{!}{
   \includegraphics[width=10cm]{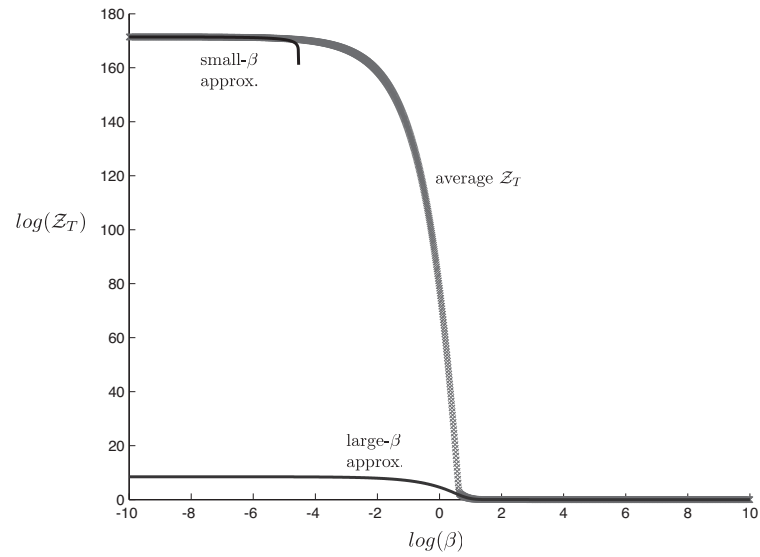} 
}   \vspace{3cm}
\caption{The average $\sZ_T$ values for different values of $\beta$, plotted on a log-log axis. The approximations for small and large $\beta$ are plotted. All values were computed by drawing $1000$ fifty taxa trees from a uniform distribution and computing normalising constants exactly using the algorithms described here.}
\end{center}
\label{fig:Zscores}
\end{figure}

\pagebreak
   \vspace*{1cm}

\begin{figure}[ht] \begin{center}
\resizebox{10.0cm}{!}{
\includegraphics[width=10cm]{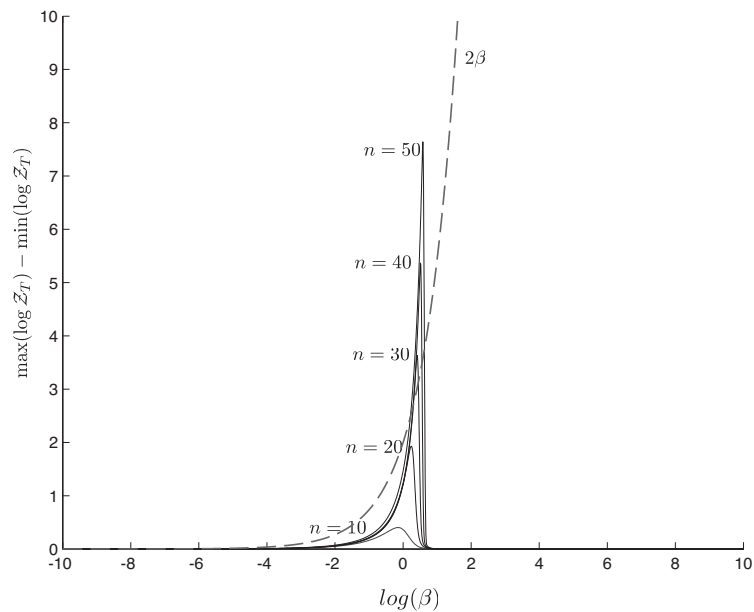}
}   \vspace{3cm}
\caption{The range of the $\sZ_T$ values computed for different $\beta$ and plotted on a log-log axis. The $\sZ_T$ values were computed by drawing $1000$ trees from a uniform distribution with $n=10,20,30,40,50$ taxa (five curves). The range is the difference between the maximum $\sZ_T$ and minimum $\sZ_T$ values, for each choice of $\beta$ and $n$. The dotted line indicates the $2 \beta$ value: when the range is less than $2 \beta$ ignoring the normalising constant has no effect on the relative order of likelihood values.}
\end{center}
\label{fig:Zrange}
\end{figure}

\pagebreak

\end{document}